\documentstyle[12pt]{article}

\catcode`\@=11
\def\marginnote#1{}
\newcount\hour
\newcount\minute
\newtoks\amorpm
\hour=\time\divide\hour by60
\minute=\time{\multiply\hour by60 \global\advance\minute
by-\hour}\edef\standardtime{{\ifnum\hour<12
\global\amorpm={am}%
        \else\global\amorpm={pm}\advance\hour by-12 \fi
        \ifnum\hour=0 \hour=12 \fi
        \number\hour:\ifnum\minute<10
0\fi\number\minute\the\amorpm}}
\edef\militarytime{\number\hour:\ifnum\minute<10
0\fi\number\minute}

\def\draftlabel#1{{\@bsphack\if@filesw {\let\thepage\relax
   \xdef\@gtempa{\write\@auxout{\string
      \newlabel{#1}{{\@currentlabel}{\thepage}}}}}\@gtempa
   \if@nobreak \ifvmode\nobreak\fi\fi\fi\@esphack}
        \gdef\@eqnlabel{#1}}
\def\@eqnlabel{}
\def\@vacuum{}
\def\draftmarginnote#1{\marginpar{\raggedright\scriptsize\tt#1}}
\def\draft{\oddsidemargin -.5truein
        \def\@oddfoot{\sl preliminary draft \hfil
        \rm\thepage\hfil\sl\today\quad\militarytime}
        \let\@evenfoot\@oddfoot \overfullrule 3pt
        \let\label=\draftlabel
        \let\marginnote=\draftmarginnote

\def\@eqnnum{(\theequation)\rlap{\kern\marginparsep\tt\@eqnlabel}%
\global\let\@eqnlabel\@vacuum}  }

\def\numberbysection{\@addtoreset{equation}{section}
        \def\theequation{\thesection.\arabic{equation}}}

\def\underline#1{\relax\ifmmode\@@underline#1\else
 $\@@underline{\hbox{#1}}$\relax\fi}

\catcode`@=12
\relax

\numberbysection

\advance \topmargin by -\headheight
\advance \topmargin by -\headsep

\evensidemargin \oddsidemargin
\def\rf#1{(\ref{#1})}
\def\lab#1{\label{#1}}
\def\nonu{\nonumber}
\def\br{\begin{eqnarray}}
\def\er{\end{eqnarray}}
\def\be{\begin{equation}}
\def\ee{\end{equation}}

\def\({\left(}
\def\){\right)}

\newcommand{\ct}[1]{\cite{#1}}
\newcommand{\bi}[1]{\bibitem{#1}}

\relax

\def\d{\delta}
\def\D{\Delta}

\def\h{{1\over 2}}

\def\pa{\partial}

\def\tp0{\Theta_{+}^{(0)}}
\def\tm0{\Theta_{-}^{(0)}}

\begin{document}
\begin{titlepage}
\begin{center}
{\large {\bf Gravitational Rainbow}}
\end{center}

\normalsize
\vskip .4in

\begin{center}

Antonio Accioly\footnote{Electronic address: accioly@ift.unesp.br} and Harold Blas\footnote{Electronic address: blas@ift.unesp.br} 

\par \vskip .1in \noindent

{\sl Instituto de F\'{\i}sica Te\'{o}rica, Universidade Estadual Paulista,}\\
{\sl Rua Pamplona 145},\\
{\sl 01405-900  S\~{a}o Paulo, SP, Brazil}\\
\end{center}

\begin{center}
\par \vskip .4in
\end{center}

 It is shown that unlike Einstein's gravity quadratic gravity produces dispersive photon propagation. The energy-dependent contribution to the deflection of photons passing by the Sun is computed and subsequently the angle at which the visible spectrum would be spread over is plotted as a function of the $R_{\mu\nu}^2-$sector mass.

\par \vskip .3in \noindent
\vspace{1 cm} 


\end{titlepage}

\vskip .3in

\newpage

\section{INTRODUCTION}

\vskip 0.2in
 \noindent According to the equivalence principle
photons follow lightlike geodesics in curved spacetime. Besides,
they are deflected in a gravitational potential by the same angle
independently of their energy or polarization. Both Einstein's
gravity \ct{r1} and $R+R^2$ gravity \ct{r2,r3} are examples
 of gravitational theories that obey this principle, which implies that
  dispersive photon propagation cannot take place within the context of
   the same. This is not true, however, as far as gravity with higher
    derivatives is concerned. Our aim here is precisely to show that
    quadratic gravity produces energy-dependent photon scattering. An
 interesting consequence of this fact is that gravity's rainbows and
 higher-derivative gravity can coexist without conflict. In this sense
 quadratic gravity is closer to quantum electrodynamics than any currently
 known gravitational theory. In fact, dispersive photon propagation is a
 trivial phenomenon in the context of QED. It is worth mentioning that
  Lafrance and Meyers \ct{r4} have shown that energy-dependent light
  scattering can also be produced within the context of a low-energy
  effective action for the electromagnetic field in curved spacetime.

On the other hand, we ought to expect a tiny value for the angle at which
the visible spectrum would be spread over. Indeed, the prediction of
general relativity for the deflection of a light ray passing close to the
Sun, namely, $1.75''$, is in good agreement with measured values for
visible light. We shall also address this question here.

We use natural units ($\hbar=c=1$) throughout. Our conventions are
$R^{\alpha}_{\beta\gamma\delta}=-\pa_{\d}\Gamma^{\alpha}_{\beta\gamma}+...,
\,\,R_{\mu\nu}=R^{\alpha}_{\mu\nu\alpha}, \,\, R=g^{\mu\nu}R_{\mu\nu}$,
where $g_{\mu\nu}$ is the metric tensor, and signature $(+---)$.

 \section{ SOLUTION TO THE LINEARIZED FIELD EQUATIONS}

 \noindent
 The theory of gravity with higher derivatives is defined
by the action \br \lab{action1} S&=&\int d^4x \sqrt{-g}\{
\frac{2R}{\kappa^2}+\frac{\alpha}{2}R^2+
\frac{\beta}{2}R^{2}_{\mu\nu}-{\cal L}_{M}\} \;\;\; , \er where
$\kappa^2=32\pi G$, with $G$ being Newton's constant, is the Einstein's
constant, $\alpha$ and $\beta$ are dimensionless parameters,  and
${\cal L}_{M}$ is the Lagrangian density for the usual matter.
This theory
 gained importance when Stelle showed that it is renormalizable,
along
 with its matter couplings \ct{r5}. In the weak field approximation ,
i.e., $g_{\mu\nu}=\eta_{\mu\nu}+\kappa  h_{\mu\nu}$, with
$\eta_{\mu\nu}=\mbox{diag}(1-1-1-1)$ and in the Teyssandier gauge \ct{r6},
namely,
 \br \lab{teyssandier} 0=\Gamma_{\mu}\equiv (1-\frac{\beta
\kappa^2}{4}\Box)
\gamma_{\mu\lambda}^{\,\,\,\,\,,\lambda}-(\alpha+\frac{\beta}{2})
\frac{\kappa^2}{2}\bar{R}_{\,,\,\mu} \;\;\; , \er with
$\gamma_{\mu\nu}\equiv h_{\mu\nu}-\h \eta_{\mu\nu}h$ and
$\bar{R}=\h \Box h-\gamma^{\mu\nu}_{\,\,\,\,\,,\mu\nu}$, the field
equations related to the action above turn out to be
\be
\lab{equations}
(1-\frac{\beta \kappa^2}{4}\Box)(-\h\Box h_{\mu\nu}+
 \frac{1}{6}\eta_{\mu\nu}\bar{R})=\frac{\kappa}{4}[T_{\mu\nu}-\frac{1}{3}
 T\eta_{\mu\nu}] \;\;\; ,
\ee where  $T_{\mu\nu}$ is the matter tensor which describes the
physical system under consideration in special relativity, i.e,
disregarding the gravitational field. The general solution of
\rf{equations} for a point particle of mass $M$ located at ${\bf
r}={\bf 0}$ is given by \ct{r7,r8,r6,r9} \br \nonu h_{00}(r)
&=&\frac{M k}{16\pi} [-\frac{1}{r} -\frac{1}{3}
\frac{e^{-m_{0}r}}{r}+ \frac{4}{3} \frac{e^{-m_{1}r}}{r}] \;\;\;,\\
h_{11}(r) &=&h_{22}(r)=h_{33}(r)\nonu\\ \nonu &=&\frac{M k}{16\pi}
[-\frac{1}{r} +\frac{1}{3} \frac{e^{-m_{0}r}}{r}+ \frac{2}{3}
\frac{e^{-m_{1}r}}{r}] \;\;\; , \er where we have assumed
$\frac{2}{k^2(3\alpha+\beta)}\equiv m_{0}^{2}>0 (3\alpha+\beta)>
0$ and $\frac{-4}{k^2 \beta}\equiv m_{1}^2>0 (-\beta>0)$, which
corresponds to the absence of tachyons (both positive and negative
 energy) in the dynamical field. Of course, the potential for quadratic
 gravity is given by the expression
\br
\lab{potential}
\kappa h_{00}(r)/2 &\equiv&V=M G[-\frac{1}{r} -\frac{1}{3}
 \frac{e^{-m_{0}r}}{r}+ \frac{4}{3} \frac{e^{-m_{1}r}}{r}]\;\;\; ,
\er which agrees asymptotically with Newton's law. At the origin
\rf{potential} tends to the finite value
$MG(\frac{m_{0}-4m_{1}}{3})$\,.

\section{ ENERGY-DEPENDENT PHOTON PROPAGATION}

 \noindent Let us now consider the scattering of a
photon by a static gravitational field generated by a localized
source like the Sun, treated as an external field. The
photon-static-external-gravitational-field vertex is given by

$$
 {\cal M}_{\mu\nu}(p,p^{\prime}) = \h \kappa\, h^{\lambda
\rho}({\bf k})
[-\eta_{\mu\nu}\eta_{\lambda\rho}p.p^{\prime}+\eta_{\lambda\rho}p_{\nu}
p^{\prime}_{\mu}+2(\eta_{\mu\nu}p_{\lambda}p^{\prime}_{\rho}-
\eta_{\nu\rho}p_{\lambda}p^{\prime}_{\mu}-
\eta{\mu\lambda}p_{\nu}p^{\prime}_{\rho}+\eta_{\mu\lambda}p.p^{\prime})]\;\;,
$$

\noindent
 where $h_{\mu\nu}({\bf k})\equiv\int d^3{\sl r}\, e^{-i{\bf
k}.{\bf r}}h_{\mu\nu}({\bf r})$ is the momentum space
gravitational field, and $p (p^{\prime})$ is the momentum of the
incoming (outgoing)  photon. Here $|{\bf p}|=|{\bf p}^{\prime}|$.
The momentum space gravitational field can be written as

\br \nonu
h_{\mu\nu}({\bf k})=h^{(E)}_{\mu\nu}({\bf
k})+h^{(R^2)}_{\mu\nu}({\bf k})+h^{(R^2_{\mu\nu})}_{\mu\nu}({\bf
k})\;\;, \er

\noindent
whereupon

\br \nonu h^{(E)}_{\mu\nu}({\bf
k})\,=\,\frac{\kappa M}{4{\bf k}^2}\eta_{\mu\nu}-\frac{\kappa
M}{2}\frac{\eta_{\mu 0}\eta_{\nu 0}}{{\bf k}^2}&,&\,\,\,
h^{(R^2)}_{\mu\nu}({\bf k})\,=\,-\frac{\kappa
M}{12}\frac{\eta_{\mu\nu}}{m_{0}^2+ {\bf k}^2}\\ \nonu
\mbox{and}\,\,\,\,\,\,\,\,\,\,\, h^{(R_{\mu\nu}^2)}_{\mu\nu}({\bf
k})&=&-\frac{\kappa M}{6}\frac{\eta_{\mu\nu}}{m_{1}^2+ {\bf
k}^2}+\frac{\kappa M}{2}\frac{\eta_{\mu 0}\eta_{\nu
0}}{m_{1}^2+{\bf k}^2}\;\;\; , \er

\noindent
 where $h^{(E)}_{\mu\nu}({\bf r})$ is
the solution of the linearized Einstein's equations supplemented
by the usual harmonic coordinate condition, namely,
$\gamma^{(E)\,\,\,\,\,,\nu}_{\mu\nu}=0$,
$\gamma_{\mu\nu}^{(E)}\equiv h_{\mu\nu}^{(E)}-\h
\eta_{\mu\nu}h^{(E)}$.

The unpolarized cross-section for the process in hand is
\br
\nonu
\frac{d \sigma}{d\Omega}= \frac{1}{32 \pi^2}\sum_{r=1}^{2}
\sum_{r^{\prime}=1}^{2} {\cal M}^{2}_{rr^{\prime}} \;\;\; ,
\er

\noindent
where ${\cal M}_{rr^{\prime}}=\epsilon^{\mu}_{r}({\bf p})
\epsilon^{\nu}_{r^{\prime}}({\bf p}^{\prime}){\cal M}_{\mu\nu}$, and
$\epsilon^{\mu}_{r}({\bf p})$ and $\epsilon^{\nu}_{r^{\prime}}
({\bf p}^{\prime})$ are the polarization vectors for the initial and
final photons, respectively. Before moving on we call attention to the
fact that the Feynman amplitude can be recast in the form
${\cal M}_{\mu\nu}={\cal M}^{(E)}_{\mu\nu}+{\cal M}^{(R^2)}_{\mu\nu}+
{\cal M}^{R_{\mu\nu}^2}_{\mu\nu},$ where
 $$
{\cal M}^{(R^2)}_{\mu\nu}=\frac{\kappa}{2}[-\frac{\kappa M}{12}
 \frac{\eta^{\lambda\rho}}{m_{0}^2+ {\bf k}^2}][-\eta_{\mu\nu}
 \eta_{\lambda\rho}p.p^{\prime}+\eta_{\lambda\rho}p_{\nu}p^{\prime}_{\mu}
+2(\eta_{\mu\nu}p_{\lambda}p^{\prime}_{\rho}-\eta_{\nu\rho}p_{\lambda}
p^{\prime}_{\mu}-\eta_{\mu\lambda}p_{\nu}
p^{\prime}_{\rho}+\eta_{\mu\lambda}\eta_{\nu\rho}p.p^{\prime})]\;\;\;.
$$

Of course, ${\cal M}_{\mu\nu}^{(R^2)} \equiv 0$, which implies
that the $R^2$ sector of the theory of gravitation with higher
derivatives does not contribute anything to the photon scattering.
After this little digression we come back to the computation of
the cross-section for the scattering of a photon by a localized
source on the basis of higher-derivative gravity. Performing the
calculation yields

\br \nonu \frac{d \sigma}{d\Omega}=
\frac{\kappa^4 E^4 M^2}{256 \pi^2} (1+\cos \phi)^2 \left[-\frac{1}{{\bf
k}^2}+\frac{1}{m_{1}^2+{\bf k}^2}\right]^2 \;\;\; ,\er

\noindent
where $E$ is the energy
of the incident photon and $\phi$ is the scattering angle. For
small angles this expression reduces to

\br \lab{section2} \frac{d
\sigma}{d\Omega}= 16 G^2 M^2 \left[-\frac{1}{\phi^2}+\frac{E^2}{m_{1}^2
+ E^2 \phi^2} \right]^2 \;\;\;.
\er

On the other hand, for small angles
\br
\lab{section3}
\frac{d \sigma}{d\Omega}= \left|\frac{r dr}{\phi d\phi}\right| \;\;\;.
\er

From \rf{section2} and \rf{section3}, we obtain at once

\br
\lab{relation} r^2 =  16 G^2
M^2 \left[\frac{1}{\phi^2}+\frac{E^2}{m_{1}^2 + E^2 \phi^2}+
\frac{2E^2}{m_{1}^2 } \ln \frac{\phi^2 E^2}{m_{1}^2 + E^2 \phi^2} \right]
\;\;\; .
\er

Thus, in the framework of higher-derivative gravity the deflection
of a photon by a localized source is a function of the energy of
the incoming photon. For a photon passing by the Sun,
\rf{relation} can be cast in the form

\br \lab{relation1}
\left(\frac{\phi}{\phi_{E}}\right)^2 -1 \,=\,\frac{1}{1+ a^2}+\frac{2}{a^2}
\ln  \frac{1}{1+ a^2} \;\;\; , \er

\noindent
where $\phi_{E}\equiv \frac{4GM}{R}$,
 with
$R$ being the Sun's radius, and $a^2\equiv \frac{m_{1}^2}{E^2
\phi^2}$. Note that the right hand side of \rf{relation1} tends to
zero as $m_{1}\rightarrow +\infty$ and, as a result, $\phi
\rightarrow \phi_{E}$ (as expected). It follows from
\rf{relation1} that for a photon just grazing the Sun's surface
$\phi$ ranges from $0^{+}$ to $1.75^{-}$ arcsec \ct{r10}.

Of course, for light rays passing close to the Sun the deflection
$\phi$ must lie in the interval $0<\phi<1.75''$. Let us show that
this is indeed the case. Consider, in this vein, the interaction
between the Sun, treated as a fixed source, and a light ray. The
associated energy-momentum tensors will be designated respectively
as $T^{\mu\nu}$ and $F^{\mu\nu}$. The current-current amplitude
for this process is given by

\br
\nonu A=g^2
T^{\mu\nu}O_{\mu\nu,\, \rho\sigma} F^{\rho\sigma}\;\;\; , \er

\noindent
where $g$
is the effective coupling constant of the theory and $O_{\mu\nu,\,
\rho\sigma}$ is the propagator for quadratic gravity. In the
Donder gauge, the propagator is given by \ct{r11}

\br \lab{prop}
O=\frac{1}{k^2}P^{1}+\frac{m^2_{1}}{k^2(m^2_{1}-k^2)}P^{2}+
\frac{m^2_{0}}{2k^2(k^2-m^2_{0})}P^{0}+\left[\frac{2}{\lambda
k^2}+ \frac{3m^2_{0}}{2k^2(k^2-m^2_{0})}\right]\bar{P}^{0}+\nonu\\
\frac{m^2_{0}}{2k^2(k^2-m^2_{0})}\bar{\bar{P}}^{0}\;\;\;, \nonu
\er

\noindent
where $P^1, P^2, P^0, \bar{P}^0$ and $\bar{\bar{P}}^0$ are the
Barnes-Rivers operators \ct{r12}, and $\lambda$ is a
gauge-parameter. But, on mass shell, $k_{\mu}T^{\mu\nu}=0$ and
$k_{\mu}F^{\mu\nu}=0$, which implies that only $P^2$ and $P^0$
will give a non null contribution
 to the current-current amplitude. Thus,
\br A=g^2
T^{\mu\nu}F^{\rho\sigma}[\frac{m^2_{1}}{k^2(m^2_{1}-k^2)}P^{2}+
\frac{m^2_{0}}{2k^2(k^2-m^2_{0})}P^{0}]_{\mu\nu\,,\,\rho\sigma}\;\;\;.
\nonu \er

Now, taking into account that \ct{r11} \br
P^2_{\mu\nu\,,\,\rho\sigma}&=&\h [\eta_{\mu\rho} \eta_{\nu\sigma}+
\eta_{\mu\sigma}
\eta_{\nu\rho}]-\frac{1}{3}\eta_{\mu\nu}\eta_{\rho\sigma}-
[P^1+\frac{2}{3}\bar{P}^0-\frac{1}{3}\bar{\bar{P}}^0]_{\mu\nu\,,\,
\rho\sigma}\,\,\,,\nonu\\
P^0_{\mu\nu\,,\,\rho\sigma}&=&\frac{1}{3} \eta_{\mu\nu}
\eta_{\rho\sigma}-
 \frac{1}{3}[\bar{P}^0+\bar{\bar{P}}^0]_{\mu\nu\,,\,\rho\sigma}\,\,\,,
 \nonu
\er and recalling that the energy-momentum tensor for light
(electromagnetic radiation) is traceless, while
$T^{\mu\nu}=\eta^{\mu 0}\eta^{\nu 0}T^{00}$ for a static source,
we promptly obtain

\br \nonu A=g^2 T^{00} F^{00}[\frac{1}{{
k}^2}-\frac{1}{{ k}^2-m_{1}^2}]\;\;\;. \er

Since $g^2 T^{00} F^{00}/k^2$ is precisely the current-current amplitude
for the interaction between the Sun and a light ray in the context of
 general relativity, we come to the conclusion that the gravitational
 deflection predicted by quadratic gravity is always smaller than that
 predicted by Einstein's theory. As a result, for light grazing the
 Sun $0<\phi<1.75''$. In a sense, this result is a major test of our
 semiclassical computation.

\section{ GRAVITATIONAL RAINBOW}

The dispersive-photon propagation we have just found represents a
tree-level violation of the equivalence principle. In addition, it tells
 us that the visible spectrum, whose wavelength ranges from $4000$
  to $7000$ (A$^{0}$), would always be spread over an angle, say
   $\D \phi$, where $\D \phi \equiv
   (\phi_{\mbox{red}}-\phi_{\mbox{violet}}$), giving rise to a
   gravitational rainbow. Such a gravity's rainbow is an intrinsic
   characteristic of quadratic gravity since the vacuum concerning
   higher-derivative gravity, like that related to QED, is a dispersive
   medium. This poses an interesting question: Is the aforementioned
   gravitational rainbow observable? Of course, we ought to expect a
   tiny value for $\D \phi$ at the Sun's limb in order not to conflict
   with well stablished results of general relativity. Indeed, Einstein's
   theory tells us that the deflection angle for a light ray passing close
   to the Sun is totally independent of the energy of the incident
   radiation and has a value equal to $1.75''-$a result that is in good
  agreement with the measured values for visible light for which there
  is perhaps $10-20\%$ uncertainty. Let us then try to answer the
     question raised above. To do that we evaluate $\D \phi$ for
     different values of $\beta$, using  \rf{relation1} and taking into
     account that experiment determines $|\beta| \leq 10^{74}$
     \ct{r13}. The result is shown in Fig. 1. A cursory inspection of
     this graph allows us to conclude that $\D \phi$ has a maximum value at
      $|\beta|=10^{64}$. For $61 \leq \mbox{Log}_{10}|\beta|\leq 69$ the
       rainbow would be in principle observable. If $|\beta| \leq 10^{60}$
       the gravity's rainbow would be practically imperceptible.
       Therefore, we come to the conclusion that in order to agree with
       the currently measured values for visible light $|\beta|
        \leq 10^{60}$.

\vskip 0.4in

\begin{center}
\noindent {\bf ACKNOWLEDGMENT}
\end{center}

\vskip 0.2in

 \noindent
 H.B. is very grateful to FAPESP for a post-doctoral fellowship.

\end{document}